\documentclass[12pt]{iopart}

\usepackage{graphicx}

\usepackage{iopams}

\begin{document}

\title[The kinematics of the charge]
{On the kinematics of the centre of charge of a spinning particle}

\author{Mart\'{\i}n Rivas}
\address{Theoretical Physics Department, The University of the Basque Country,\\ 
Apdo.~644, 48080 Bilbao, Spain}
\ead{martin.rivas@ehu.es}

\begin{abstract}
In particle physics, most of the classical models 
consider that the centre of mass and centre of charge of an elementary particle, 
are the same point.
This presumes some particular relationship between the charge and mass distribution, 
a feature which cannot be checked experimentally. 
In this paper we give three different kinds of arguments suggesting 
that, if assumed different points, the centre of charge of an elementary spinning 
particle moves in a helical motion at the speed of light, and it thus satisfies,
in general, a fourth order differential equation. If assumed a kind of rigid body structure,
it is sufficient the description of the centre of charge to describe also 
the evolution of the centre of mass
and the rotation of the body.
This assumption of a separation betwen the centre of mass and centre of
charge gives a contribution to the spin of the system 
and also justifies the existence of a magnetic
moment produced by the relative motion of the centre of charge.
This corresponds to an improved model of a charged elementary particle,
than the point particle case.
This means that a Lagrangian formalism for describing elementary spinning
particles has to depend, at least, up to the acceleration of the position of the charge, 
to properly obtain fourth order dynamical equations. This result is compared with the description
of a classical Dirac particle obtained from a general Lagrangian 
formalism for describing spinning particles.
\end{abstract} 


\maketitle

\section{Introduction}\label{sec:intro}

It is implicitely assumed that the independent degrees of freedom of any mechanical system satisfy, in general,
second order differential equations, as it happens for the Newton equations of the centre of mass of the mechanical system. 
This is reflected in the general assumption that the Lagrangian of any mechanical system is an explicit 
function $L(t,q_i,\dot{q}_i)$ of the time $t$, the $n$ degrees of freedom $q_i$ and their time derivatives $\dot{q}_i$.
Lagrangians depending on higher order derivatives are scarcely used, and their usefulness is left to some
especific problems. In this work we are going to give three different kinds of 
arguments to justify the use of Lagrangians
depending up to the acceleration of the centre of charge for describing charged
elementary spinning particles. The centre of charge of an elementary particle 
thus satisfies fourth order differential equations.

In section \ref{rigid} we analyze the motion of the centre of charge of a charged rigid
body, considered as a model of a classical elementary particle. In section \ref{invariance}
we suggest that the order of the invariant differential equation satisfied by the evolution
of a point depends on the number of parameters of the kinematical group of the theory. Since 
the Galilei and Poincar\'e groups are ten-parameter groups, and if the point represents the position of the centre 
of charge of an elementary particle, the corresponding differential
equation must be of fourth order. Section \ref{geometrical} analyzes the most general differential
equation satisfied by a point in three dimensional space, which is of fourth order. 
Finally, section \ref{Dirac} describes the model of a Dirac particle obtained from a general
formalism for describing classical spinning particles and compares its structure with the previous
predicted motions.

\section{Rigid body arguments}
\label{rigid}

Let us consider that an elementary particle is described as a nonrelativistic rigid body.
A rigid body is a mechanical system of six degrees of freedom. Three represent the position
of a point and the other three the orientation of a body frame attached to that point.
Usually, it is described by the location of the centre of mass, which is represented by the point ${\bi q}$,
and the orientation by the principal axis of inertia located around ${\bi q}$.
The centre of mass satisfies second order dynamical equations and moves like a point of
mass $m$, the total mass of the system, under the total external force. The torques of
the external forces produce a change in the orientation of the body. In this way a rigid body
moves and rotates. 

\begin{figure}
\begin{center}\includegraphics[width=7cm]{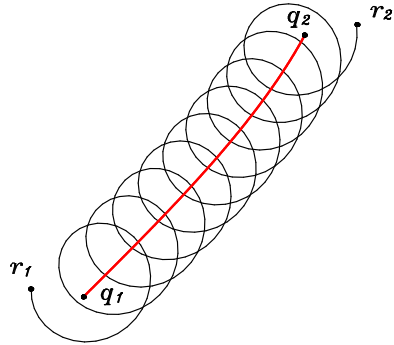}\end{center}
\caption{Motion of the centre of mass ${\bi q}$ and and another arbitrary point ${\bi r}$
of a rigid body.}\label{fig:1}
\end{figure}
If we describe the evolution of a different point ${\bi r}$,
it will follow, in general, a helical trajectory around the centre
of mass, like the one depicted in the figure~\ref{fig:1}.

If an elementary particle is a charged rigid body, it is clear that we also need to know its electromagnetic
structure, which can be reduced to the knowledge of the centre of charge and the different multipoles.
If assumed some spherical symmetry for the electric field produced by the particle,
we are left with the location of the centre of charge and no further multipoles. The position of 
this point will be used to determine there the actions of the external fields and to compute from there the
fields generated by the particle.
In general, depending how the mass and charge are distributed, these two points will be different
points as we shall assume here. Therefore, if we try to describe the evolution of the centre of mass,
we have to determine also at any time the location of the centre of charge to compute the external forces.
Newton's dynamical equations for the centre of mass will be written as
 \begin{equation}
m\frac{d^2{\bi q}}{dt^2}=e\left({\bi E}(t,{\bi r})+\frac{d{\bi r}}{dt}\times{\bi B}(t,{\bi r})\right)={\bi F}(t,{\bi r},d{\bi r}/dt).
 \label{eq:cm}
 \end{equation}
The electromagnetic force ${\bi F}$ depends, in general, on the electric and magnetic external
fields defined at the charge position ${\bi r}$ and on the velocity
of the charge $d{\bi r}/dt$ which appears in the magnetic term.

\begin{figure}\begin{center}\includegraphics[width=7cm]{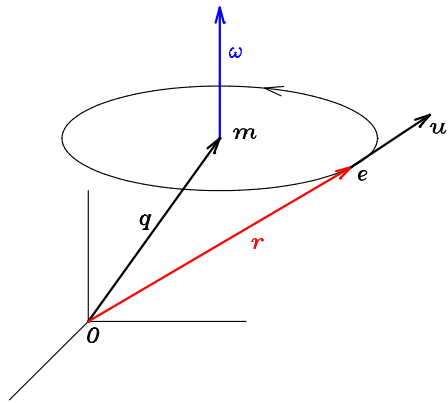}\end{center}
\caption{Motion of the centre of charge ${\bi r}$ around the centre of mass at rest ${\bi q}$.
}\label{fig:2}
\end{figure}

The relative motion between ${\bi r}$ and ${\bi q}$ when analyzed by an observer who sees the centre of 
mass at rest like in figure \ref{fig:2} is a circular motion with constant velocity, in particular in the free case.
If we define a unit vector ${\bi n}$ in the direction of the normal acceleration $d^2{\bi r}/dt^2$ 
of point ${\bi r}$, 
\[
{\bi n}=\frac{1}{\omega^2R}\frac{d^2{\bi r}}{dt^2},\quad 
\]
where $R$ is the radius of the circular motion,
then, the centre of mass position can be written as
 \begin{equation}
{\bi q}(t)={\bi r}(t)+\frac{1}{\omega^2}\frac{d^2{\bi r}}{dt^2}.
 \label{eq:q}
 \end{equation}
For the centre of mass observer, the angular velocity
is also orthogonal to the plane subtended by 
the velocity ${\bi u}=d{\bi r}/dt$ and acceleration $d{\bi u}/dt$ of point ${\bi r}$, and given by
 \begin{equation}
\bomega=\frac{1}{u^2}\,{\bi u}\times\frac{d{\bi u}}{dt}.
 \label{eq:w}
 \end{equation}
In this way, we have also solved the problem of the rotation of the charged rigid body by analyzing the evolution
of just the centre of charge. 

Therefore, it will be simpler, from a theoretical point of view,
just to describe the evolution of a single point, the centre of charge ${\bi r}$, 
instead of the orientation of the rigid body and its centre of mass ${\bi q}$, 
which will be in some average position of point ${\bi r}$,
and obtained from (\ref{eq:q}) once the trajectory of ${\bi r}$ is computed.
The elimination of the ${d^2{\bi q}}/{dt^2}$ among equations (\ref{eq:cm}) and (\ref{eq:q}) will give 
us, in general, a fourth order differential equation for the variable ${\bi r}$. 
If the rigid body is no longer free, we shall assume that its stiffness means that
the relationship between ${\bi q}$, $\bomega$ and ${\bi r}$ and their derivatives given in (\ref{eq:q}) 
and (\ref{eq:w}) still remain valid, even in the presence of an external interaction.

Therefore, if the centre of mass and centre of charge of a rigid body are different
points, it is sufficient to describe the evolution of the centre of charge. Then, the rigid body is reduced
to a system of only three degrees of freedom which satisfy fourth-order differential equations. 
We see that a Lagrangian depending on the acceleration of point ${\bi r}$, could reproduce such equations.
The following nonrelativistic free Lagrangian
\[
L=\frac{m}{2}\left(\frac{d{\bi r}}{dt}\right)^2-\frac{m}{2\omega^2}\left(\frac{d^2{\bi r}}{dt^2}\right)^2
\]
produces the dynamical equations for the point ${\bi r}$
\[
\frac{d^4{\bi r}}{dt^4}+\frac{1}{\omega^2}\frac{d^2{\bi r}}{dt^2}=0,
\]
which can be factorized in terms of a point ${\bi q}$, defined as in (\ref{eq:q}) in the form
\[
\frac{d^2{\bi q}}{dt^2}=0,\quad \frac{d^2{\bi r}}{dt^2}+\omega^2({\bi r}-{\bi q})=0,
\]
a free motion for the point ${\bi q}$ which can be interpreted as the centre of mass, and a harmonic motion
of point ${\bi r}$ around ${\bi q}$ with a constant frequency $\omega$. It is the presence of the acceleration
term in the Lagrangian which produces the definition of a point ${\bi q}$, 
different from ${\bi r}$, moving freely. This Lagrangian describes a particle with spin \cite{Rivasbook}, where
its detailed analysis can be found in this reference.
\section{Invariance arguments}
\label{invariance}
Let us consider the trajectory ${\bi r}(t)$, $t\in[t_1,t_2]$ followed by a point of a mechanical 
system for an arbitrary inertial observer $O$. Any other inertial observer $O'$ 
is related to the previous one by a transformation
of the spacetime kinematical group such that their relative spacetime measurements of any 
spacetime event are given by
\[
t'=T(t,{\bi r}; g_1,\ldots,g_\alpha),\quad {\bi r}'={\bi R}(t,{\bi r}; g_1,\ldots,g_\alpha),
\]
where the functions $T$ and ${\bi R}$ define the corresponding spacetime transformation 
of the kinematical group $G$, of parameters $(g_1,\ldots,g_\alpha)$, among any two observers. 
Then the description of the trajectory of that point 
for observer $O'$ is obtained from
\[
t'(t)=T(t,{\bi r}(t); g_1,\ldots,g_\alpha),\quad {\bi r}'(t)={\bi R}(t,{\bi r}(t); g_1,\ldots,g_\alpha),\quad \forall t\in[t_1,t_2].
\]
If we eliminate $t$ as a function of $t'$ from the first equation and substitute into the second 
we shall get
\begin{equation}
{\bi r}'(t')={\bi r}'(t'; g_1,\ldots,g_\alpha).
 \label{eq:robs}
 \end{equation}
Since observer $O'$ is arbitrary, equation (\ref{eq:robs}) represents the complete set of 
trajectories of the point for all inertial observers. 
Elimination of the $\alpha$ group parameters among the function ${\bi r}'(t')$
and their time derivatives will give us the differential equation satisfied by all the trajectories 
of the point. 
Let us assume that the trajectory is unrestricted
in such a way that the above group parameters are essential in the sense that no smaller number of them
gives the same family of trajectories.
This differential equation is invariant by construction because it is independent
of the group parameters and therefore independent of any inertial observer.
If $G$ is either the Galilei or Poincar\'e group,
it is a ten-parameter group so that we have to work out in general up to the fourth derivative 
to obtain sufficient equations to eliminate the group parameters. 
Therefore the order of the differential equation is dictated by the
number of parameters and the structure of the kinematical group. If the point ${\bi r}$ 
represents the position of the centre of charge of an elementary particle we get 
again that it satisfies, in general, a fourth order differential equation.
But at the same time it is telling us that to obtain the invariant differential equation satisfied
by the centre of charge of an elementary particle, it is sufficient to obtain its trajectory in an arbitrary
reference frame, and to follow the above procedure of elimination of the group parameters.
\section{Geometrical arguments}
\label{geometrical}

A continuous and differentiable curve in three-dimensional space, 
${\bi r}(s)$, is a regular curve if at any point $s$ has a tangent vector $\dot{\bi r}(s)\neq0$. 
It has associated three orthogonal unit vectors, ${\bi t}$,
${\bi n}$ and ${\bi b}$, called respectively the tangent, normal and binormal. 
If using the arc length $s$ as the curve parameter, they satisfy the Frenet-Serret equations
\[
\dot{\bi t}=\kappa{\bi n},\quad \dot{\bi n}=-\kappa{\bi t}+\tau{\bi b},\quad
\dot{\bi b}=-\tau{\bi n},
\] 
where the overdot means the derivative with respect to $s$. 
The knowledge of the curvature $\kappa(s)$ and torsion $\tau(s)$, 
together the boundary values ${\bi r}(0)$, ${\bi t}(0)$,
${\bi n}(0)$ and ${\bi b}(0)={\bi t}(0)\times{\bi n}(0)$, completely
determine the curve, because the above equations are integrable. In terms of the 
vector $\bomega=\tau{\bi t}+\kappa{\bi b}$, known as Darboux vector,
the Frenet-Serret equations can be rewritten as
\[
\dot{\bi t}=\bomega\times{\bi t},\quad\dot{\bi n}=\bomega\times{\bi n},\quad\dot{\bi b}=\bomega\times{\bi b},
\]
so that, in units of arc length, Darboux vector represents the instantaneous 
angular velocity undergone by the local triad frame.

By changing the notation ${\bi r}^{(k)}(s)\equiv d^k{\bi r}/ds^k$, we have  
\[
{\bi r}^{(1)}={\bi t},\quad {\bi r}^{(2)}=\kappa{\bi n},\quad {\bi r}^{(3)}=\dot{\kappa}{\bi n}+\kappa(-\kappa{\bi t}+\tau{\bi b})
\]
and this allows us to eliminate the three unit vectors ${\bi t}$,
${\bi n}$ and ${\bi b}$, in terms of the three derivatives ${\bi r}^{(k)}$, $k=1,2,3$. If we 
replace them in the next order derivative ${\bi r}^{(4)}$, one obtains the most general 
differential equation satisfied by the point ${\bi r}$, i.e.,
the fourth order differential system
 \begin{equation}
{\bi r}^{(4)}-\frac{2\dot{\kappa}\tau+\dot{\tau}\kappa}{\kappa\tau}{\bi r}^{(3)}
+\left(\kappa^2+\tau^2+\frac{\dot{\kappa}\dot{\tau}-\tau\ddot{\kappa}}{\kappa\tau}+\frac{2\dot{\kappa}^2
}{\kappa^2}\right){\bi r}^{(2)}
+\frac{\kappa}{\tau}(\dot{\kappa}{\tau}-\dot{\tau}\kappa){\bi r}^{(1)}=0.
 \label{eq:masgeneral}
 \end{equation}

Let us consider that an elementary particle, instead of being a rigid body, is just 
a localized mechanical system. By localized we mean that,
at least, it is described by the evolution of a single point ${\bi r}$. 
This point could be the centre of mass, but, as mentioned before, 
in order to determine the external forces, we also need
to know the location of the centre of charge to compute the actions of the external fields. 
Let us assume that the elementary particle is charged. 
By the previous arguments,
if its electric field is spherically symmetric, we are reduced
to know the evolution just of the centre of charge. 
The particle will have a centre of mass but 
we make the assumption that the centre of mass
and the centre of charge are not necessarily the same point. 

Let us consider that the geometrical regular curve ${\bi r}(s)$ in three-dimensional space represents
the trajectory of the centre of charge of an elementary particle. When the corresponding inertial 
observer uses its time as the evolution parameter, kinematics enters into the scene.
Let us assume now that the motion of the particle is free. 
This means that we cannot distinguish one point of the evolution from another, so that the motion
has to be at a constant velocity such that
the arc length $ds=|{\bi u}|dt$, where ${\bi u}=d{\bi r}/dt$ is the velocity
of the charge, must be independent of the time $t$. 
Otherwise, if $ds$ is not the same we can distinguish one instant of the evolution from another,
as far as the displacement of the charge is concerned.
At the same time, Darboux vector has to be also independent of time.
The Frenet-Serret triad moves and rotates in a free motion with constant linear and angular velocities.
The curvature and torsion are necessarily constants of the motion.

Thus $\dot{\kappa}=\dot{\tau}=0$, and, in the free case, the equations (\ref{eq:masgeneral})
are reduced to
\[
{\bi r}^{(4)}+\left(\kappa^2+\tau^2\right){\bi r}^{(2)}=\frac{d^2}{ds^2}\left({\bi r}^{(2)}+\left(\kappa^2+\tau^2\right){\bi r}\right)=0.
\] 
If the curvature and torsion are constant the curve is a helix, 
which can be factorized in terms of a
central point 
\[
{\bi q}={\bi r}+\frac{1}{\omega^2}\frac{d^2{\bi r}}{dt^2}, \quad \frac{d^2{\bi q}}{ds^2}=0,\quad \omega^2=\kappa^2+\tau^2
\]
which is moving in a straight
trajectory, while the point ${\bi r}$ satisfies
\[
\frac{d^2{\bi r}}{dt^2}+\omega^2\,({\bi r}-{\bi q})=0,
\]
an isotropic harmonic motion of frequency $\omega$,
around point ${\bi q}$. 
The point ${\bi q}$ clearly represents the centre of mass position of the free particle and its expression
in terms of the centre of charge position is exactly the same as in the case for the rigid body (\ref{eq:q}).

The centre of charge of a free elementary
particle is describing a helix at a constant velocity for any inertial observer.
If we make a nonrelativistic analysis, the relationship of the velocity measurements among two arbitrary
inertial observers $O$ and $O'$, is given by ${\bi u}'=R{\bi u}+{\bi v}$, where ${\bi v}$ is the constant 
velocity of $O$ as measured by $O'$ and the constant rotation matrix $R$ is their relative orientation.
Now,
 \begin{equation}
{u'}^2=u^2+v^2+2{\bi v}\cdot R{\bi u}.
 \label{una}
 \end{equation}
In a relativistic analysis
\[
{\bi u}'=\frac{R{\bi u}+\gamma{\bi v}+\frac{\gamma^2}{(1+\gamma)c^2}({\bi v}\cdot R{\bi u}){\bi v}}{\gamma(1+{\bi v}\cdot R{\bi u}/c^2)},\quad \gamma=(1-v^2/c^2)^{-1/2},
\]
where ${\bi v}$ is also the velocity of observer $O$ measured by $O'$ and $R$ represents
their relative orientation, and thus
 \begin{equation}
u^{'2}=\frac{u^2-c^2}{\gamma^2\left(1+{\bi v}\cdot R{\bi u}/c^2\right)^2}+c^2.
 \label{dos}
 \end{equation}
Taking the time derivative of both expressions (\ref{una}) and (\ref{dos})
if $u'$ has to be also constant for observer $O'$, we get that ${\bi v}\cdot R\dot{\bi u}=0$, 
irrespective of ${\bi v}$ and of the rotation matrix $R$.
This means that the vector ${\bi u}$ must be a constant vector. The centre of charge necessarily moves along
a straight trajectory at a constant velocity, for every inertial observer, and the above general
helix degenerates into a straight line and because the point is not accelerated ${\bi q}={\bi r}$.

This is the usual description of the spinless or pointlike free
elementary charged particle, whose centre of charge and centre of mass are 
represented by the same point.

However, in this relativistic analysis, there is one alternative not included in the nonrelativistic
approach. The possibility that the charge of an elementary particle will be moving at the speed of light
and, in that case, $u=u'=c$, for any inertial observer. 
This means that the centre of the helix is always moving at a velocity $|d{\bi q}/dt|<c$, and, 
if it represents the centre of mass, this particle is a massive particle. 
In a variational description of this system
the Lagrangian should depend up to the acceleration of the point ${\bi r}$ 
in order to obtain fourth order differential equations. This dependence on the acceleration 
will give a contribution to the spin of the particle and there is also another contribution
from the rotation of the system, because the body frame 
rotates with angular velocity $\bomega$. The motion of the charge around the centre of mass 
produces the magnetic moment of the particle.

When we introduce the time as the curve parameter if the trajectory is necessarily a helix at a 
constant velocity $u$, it is equivalent the use of the arc length $s=ut$ or the time as the curve parameter.
But if the curve is regular it must have at any point a tangent vector ${\bi r}^{(1)}\neq0$.
This implies that the velocity of the charge can never be reached by any inertial observer. 
Otherwise, the velocity of the charge will be zero at a certain time but different from zero for subsequent times,
because the point is accelerated, which is contradictory
with the requirement that the point describes a trajectory with a constant velocity.

This kinematical argument requires that the kinematical group of spacetime symmetries must contain a limit
velocity which cannot be reached by any inertial observer. Among the possible ten-dimensional kinematical
groups found by Bacry and Levy-Leblond \cite{BLL}, only the De Sitter groups $SO(3,2)$ and $SO(4,1)$ and the 
Poincar\'e group ${\cal P}$ contain such a limit velocity $c$. If the spacetime is flat then only the Poincar\'e kinematical
description is singled out by this requirement, so that the charge, necessarily, 
must be moving at this limit velocity.

In summary, there are only two possibilities for a free motion of the centre of 
charge of an elementary particle.
One, the charge is moving along a straight line at any constant velocity, the centre of mass is the same point, 
the system has no magnetic moment
and it can be described either in a relativistic or a non-relativistic framework.
In the other, the particle has spin and magnetic moment, and the charge moves along a helix at the speed
of light. Because all known elementary particles, quarks and leptons, are spin $1/2$ particles, 
and the charged ones have magnetic moment, we are left only with this last possibility if we want to give
a more improved description of an elementary particle. Here, only the relativistic description is allowed.

This is consistent with Dirac's theory of the electron, 
because the eigenvalues of the components of Dirac's velocity operator are $\pm c$. 
This means that Dirac's spinor $\psi(t,{\bi r})$ is expressed in terms of the 
position of the charge ${\bi r}$, because the external fields
$A_\mu(t,{\bi r})$ are defined and computed at this point.

This last possiblity is the description of the centre of charge of a relativistic spinning 
elementary particle obtained in the kinematical formalism \cite{Rivasbook}, 
and which satisfies Dirac's equation when quantized. The classical structure of this Dirac particle is analyzed
in the next section.

In this formalism Dirac particles are localized and also orientable mechanical systems. By orientable
we mean that we have to attach to the above point ${\bi r}$,
a local cartesian frame to describe its spatial orientation. This frame could be the Frenet-Serret triad. 
The rotation of the frame will also contribute
to the total spin of the particle. When quantizing the system, 
the spin $1/2$ is coming from the presence of the orientation variables.
Otherwise, if there are no orientation variables, no spin $1/2$ structure is described 
when quantizing the system. This twofold structure of the classical spin has produced a
pure kinematical interpretation of the gyromagnetic ratio \cite{g2}.
The dependence of the Lagrangian on the acceleration is necessary for the particle to have magnetic moment
and for the separation between the centre of mass and centre of charge.

\section{Kinematical description of a Dirac particle}
\label{Dirac}
In the kinematical formalism \cite{Rivasbook}, an elementary particle is, by definition, 
a mechanical system which in addition to being indivisible, as a consequence of the atomic hypothesis  \cite{atomic}, 
it can never be deformed so that all allowed states
are only kinematical modifications of any one of them. This means that when the state
of an elementary particle changes it is possible to find an inertial observer
who measures the particle in the same state as before. An electron, if not annihilated, 
always remains an electron under any external force. This means that in a variational 
approach the initial $x_1$ and final $x_2$ states of the evolution are related by a transformation of the
kinematical group $x_2=gx_1$. Therefore, the boundary variables
of the variational approach, necessarily span a homogeneous space of the kinematical 
group of space-time symmetries. When quantizing all classical systems 
characterized by such homogeneous spaces, their Hilbert space of pure states carries a projective
unitary irreducible representation of the kinematical group \cite{RivasQ}. It thus satisfies Wigner's
definition of a quantum elementary particle.

In this way, the parameters of the kinematical group become the classical variables we need to consider,
as the boundary values of the variational formalism for describing an elementary particle. 
In the relativistic and non-relativistic approach, these
variables are reduced to the ten variables $t,{\bi r},{\bi u}$ and $\balpha$, interpreted respectively 
as the time, position of the charge, velocity of the charge and orientation.
In the relativistic case we have three
disjoint, maximal homogeneous spaces of the Poincar\'e group 
spanned by these variables with the constraint either $u<c$, $u=c$
or $u>c$. It is the manifold with $u=c$, as suggested by the kinematical arguments of the previous section, 
which leads to Dirac's equation when quantizing the system. Because
the Lagrangian depends on the next order derivative of the boundary variables, it thus depends also on the 
acceleration of the point ${\bi r}$ and on the angular velocity. It is thus clear that the point ${\bi r}$
cannot be the centre of mass because satisfies fourth-order differential equations. Because the external
interaction is defined at this point ${\bi r}$, is why we consider it represents the position of the charge.

Then, for a Dirac particle, the charge located at point ${\bi r}$,
is moving at the speed of light $u=c$. 
The classical expression which gives rise to Dirac's equation is 
\[
H={\bi P}\cdot{\bi u}+\frac{1}{c^2}{\bi S}\cdot\left(\frac{d{\bi u}}{dt}\times{\bi u}\right),
\]
where the energy $H$ is expressed as the sum of two terms, ${\bi P}\cdot{\bi u}$, or translational energy
and the other, which depends on the spin of the system, or rotational energy. 
The spin comes from the dependence of the Lagrangian $L$
of both, the acceleration $\dot{\bi u}$, and the angular velocity  ${\bomega}$, and if we define
\[
{\bi U}=\frac{\partial L}{\partial \dot{\bi u}},\quad {\bi W}=\frac{\partial L}{\partial\bomega},
\] 
it takes the form
\[
{\bi S}={\bi u}\times{\bi U}+{\bi W}={\bi Z}+{\bi W}.
\]
The first part ${\bi Z}={\bi u}\times{\bi U}$, or {\it zitterbewegung} part, 
is related to the separation between the centre of charge from the centre of mass and 
takes into account this relative orbital motion. It quantizes with integer values. 
The second part ${\bi W}$
is the rotational part of the body frame and quantizes with both integer and half-integer values.
The total angular momentum with respect to the origin of observer's frame is
\[
{\bi J}={\bi r}\times{\bi P}+{\bi S},
\]
so that the spin ${\bi S}$ is the angular momentum of the system with respect 
to the centre of charge ${\bi r}$,
and not with respect to the centre of mass ${\bi q}$. By this reason, it
is not a conserved quantity for a free particle,
but satisfies the dynamical equation
 \begin{equation}
\frac{d{\bi S}}{dt}={\bi P}\times{\bi u}.
 \label{eq:dynspin}
 \end{equation}
This is exactly the same dynamical equation satisfied by Dirac's spin operator in the quantum case.
This has to be taken into account when comparing the analysis of this spin 
with other approaches, for instance, with Bargmann-Michel-Telegdi spin observable \cite{BMT}, which clearly represents the
angular momentum with respect to the centre of mass of the system. 
Once a mechanical system has two distinguished points, the centre of charge and centre of mass, 
we must clarify with respect to which of these points is defined the angular momentum of the system.
This is important, for instance, in the so called
{\it proton spin crisis}.  If the spin of the proton is the angular momentum with respect to its 
centre of mass, and we add the three Dirac spin operators of the three quarks we cannot obtain
the spin of the proton.
What we have to add are the three spins of the quarks with respect to their corresponding centre of mass,
if assumed that the motion of the quarks is in an $l=0$ orbital angular momentum state.

\begin{figure}
\begin{center}\includegraphics[width=7cm]{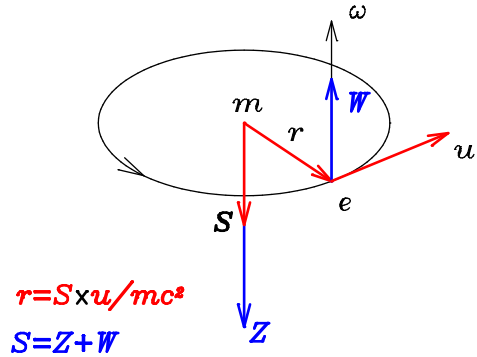}\end{center}
\caption{\label{fig:3}{Motion of the charge of the electron in the centre of mass frame. The magnetic moment
of the particle is produced by the motion of the charge. The total spin
${\bi S}$ is half the value of the zitterbewegung part ${\bi Z}$ when quantizing the sytem, so that when expressing the magnetic
moment in terms of the total spin we get a $g=2$ gyromagnetic ratio \cite{g2}. 
The body frame attached to the point ${\bi r}$, which could be Frenet-Serret triad,
rotates with angular velocity $\bomega$, has not been depicted.}} 
\end{figure}

When expressed Dirac's spin and the centre of mass position 
in terms of the velocity and acceleration of the charge they take, respectively, the form
\[
{\bi S}=\left(\frac{H-{\bi u}\cdot{\bi P}}{\left({d{\bi u}}/{dt}\right)^2}\right)\,\frac{d{\bi u}}{dt}\times{\bi u},\quad 
{\bi q}={\bi r}+\frac{c^2}{H}\left(\frac{H-{\bi u}\cdot{\bi P}}{(d{\bi u}/dt)^2}\right)\frac{d{\bi u}}{dt}
\]
Dirac's spin is always orthogonal to the osculator plane of the trajectory 
of the charge ${\bi r}$, 
in the direction opposite to the binormal for a positive energy particle, and in the 
opposite direction for the antiparticle. This implies a difference in chirality between matter and 
antimatter. 
The acceleration of the charge is pointing from ${\bi r}$ to the centre of mass, as it corresponds to 
a helix.
It is shown that the dynamical equation of point ${\bi r}$ for the free particle and 
in the centre of mass frame is given by
 \begin{equation}
{\bi r}=\frac{1}{mc^2}{\bi S}\times{\bi u},
 \label{eq:dyq}
 \end{equation}
and where the spin vector ${\bi S}$ is constant in this frame, as depicted in Fig.~\ref{fig:3}.
The radius of the zitterbewegung motion is $R=S/mc$, and the angular velocity $\omega=mc^2/S$. 
When considered in the centre-of mass frame it is a system of three degrees of freedom; two are the 
$x$ and $y$ components of the position of the charge on the zitterbewegung plane and the third is the phase
of the rotation of the body frame. This phase is the same as the phase of the orbital motion and because
the velocity $u=c$ is constant, we are just left with a single and independent degree of freedom, for instance,
the $x$ coordinate. The Dirac particle, when considered in the centre of mass frame, is equivalent to 
a one-dimensional harmonic oscillator of frequency $\omega$. 
If we quantize this harmonic oscillator, because the centre of mass is at rest 
the oscillator is in its ground state of energy
$\hbar\omega/2=mc^2$. No further excited states are allowed according to the atomic principle \cite{atomic}.
The classical spin parameter $S$ thus becomes $\hbar/2$, when quantized.

\begin{figure}
\begin{center}\includegraphics[width=7cm]{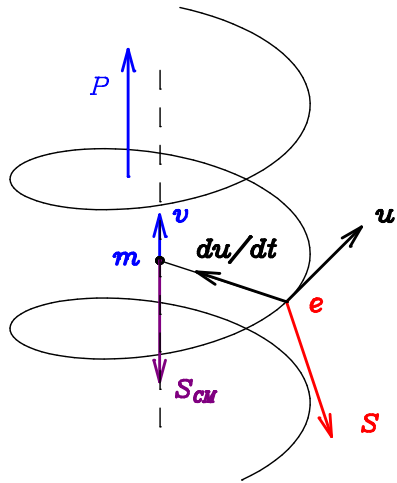}\end{center}
\caption{\label{fig:4}{Precession of Dirac's spin ${\bi S}$ along the linear momentum ${\bi P}$. The tranversal
motion of the charge takes a time $\gamma(v)$ longer than when the centre of mass at rest, 
to complete a turn. 
The three vectors ${\bi u}$, $d{\bi u}/dt$
and $-{\bi S}$, properly normalized, form the Frenet-Serret triad of the motion of the charge. 
The spin with
respect to the centre of mass ${\bi S}_{CM}$, is a constant of the motion for the free particle.}} 
\end{figure}

When seen from an arbitrary observer (see Figure~\ref{fig:4}), the motion of the charge is a helix, 
so that according to (\ref{eq:dynspin}) Dirac's spin precess around the direction of the 
conserved linear momentum ${\bi P}$. The spin with respect to the centre of mass is defined as
\[
{\bi S}_{CM}={\bi S}+({\bi r}-{\bi q})\times{\bi P}.
\]
It is a conserved quantity for a free particle. 
The centre of mass velocity is ${\bi v}=d{\bi q}/dt$, 
and the linear momentum is written as usual
as ${\bi P}=\gamma(v)m{\bi v}$. This means that the transversal motion of the charge
is at the velocity $\sqrt{c^2-v^2}$. A moving electron takes a time $\gamma(v)$ times longer
than for an electron at rest to complete a turn, as a result of the time dilation measurement.
The faster the centre of mass of the electron moves the slower is the rotation frequency 
of the centre of charge around the centre of mass. The internal clock of a fast electron is running
slower.

\ack{
I thank my colleague J M Aguirregabiria for the use of his excellent
Dynamics Solver program \cite{JMA} with which the numerical computations of the electron trajectories
have been done.
This work was supported by The University of the Basque Country
(Research Grant~9/UPV00172.310-14456/2002).}


\end{document}